\catcode`\@=11                                   % To make protected \def's

%************************************************************
%*
%*            Font set-up
%*
%************************************************************

%************** 5-point fonts *******************************

\font\fiverm=cmr5                         % roman
\font\fivemi=cmmi5                        % math italic
\font\fivesy=cmsy5                        % math symbols
\font\fivebf=cmbx5                        % bold face

\skewchar\fivemi='177
\skewchar\fivesy='60

%************** 6-point fonts *******************************

\font\sixrm=cmr6                          % roman
\font\sixi=cmmi6                          % math italic
\font\sixsy=cmsy6                         % math symbols
\font\sixbf=cmbx6                         % bold face

\skewchar\sixi='177
\skewchar\sixsy='60

%************** 7-point fonts *******************************

\font\sevenrm=cmr7                        % roman
\font\seveni=cmmi7                        % math italic
\font\sevensy=cmsy7                       % math symbols
\font\sevenit=cmti7                       % italic
\font\sevenbf=cmbx7                       % bold face

\skewchar\seveni='177
\skewchar\sevensy='60

%************** 8-point fonts *******************************

\font\eightrm=cmr8                        % roman
\font\eighti=cmmi8                        % math italic
\font\eightsy=cmsy8                       % math symbols
\font\eightit=cmti8                       % italic
                       % slanted
\font\eightbf=cmbx8                       % bold face
                       % typewriter
                       % sans serif

\skewchar\eighti='177
\skewchar\eightsy='60

%************** 9-point fonts *******************************

\font\ninei=cmmi9
\font\ninesy=cmsy9

\skewchar\ninei='177
\skewchar\ninesy='60

%************** 10-point fonts ******************************

\font\tenrm=cmr10                         % roman
\font\teni=cmmi10                         % math italic
\font\tensy=cmsy10                        % math symbols
\font\tenex=cmex10                        % math extension
\font\tenit=cmti10                        % italic
\font\tensl=cmsl10                        % slanted
\font\tenbf=cmbx10                        % bold face
\font\tentt=cmtt10                        % typewriter
\font\tenss=cmss10                        % sans serif
\font\tensc=cmcsc10                       % small caps
\font\tenbi=cmmib10                       % bold math

\skewchar\teni='177
\skewchar\tenbi='177
\skewchar\tensy='60

\def\tenpoint{\ifmmode\err@badsizechange\else
	\textfont0=\tenrm \scriptfont0=\sevenrm \scriptscriptfont0=\fiverm
	\textfont1=\teni  \scriptfont1=\seveni  \scriptscriptfont1=\fivemi
	\textfont2=\tensy \scriptfont2=\sevensy \scriptscriptfont2=\fivesy
	\textfont3=\tenex \scriptfont3=\tenex   \scriptscriptfont3=\tenex
	\textfont4=\tenit \scriptfont4=\sevenit \scriptscriptfont4=\sevenit
	\textfont5=\tensl
	\textfont6=\tenbf \scriptfont6=\sevenbf \scriptscriptfont6=\fivebf
	\textfont7=\tentt
	\textfont8=\tenbi \scriptfont8=\seveni  \scriptscriptfont8=\fivemi
	\def\rm{\tenrm\fam=0 }%
	\def\it{\tenit\fam=4 }%
	\def\sl{\tensl\fam=5 }%
	\def\bf{\tenbf\fam=6 }%
	\def\tt{\tentt\fam=7 }%
	\def\ss{\tenss}%
	\def\sc{\tensc}%
	\def\bmit{\fam=8 }%
	\rm\setparameters\setbaselines\fi}

%************** 12-point fonts ******************************

\font\twelverm=cmr12                      % roman
\font\twelvei=cmmi12                      % math italic
\font\twelvesy=cmsy10       scaled\magstep1             % math symbols
\font\twelveex=cmex10       scaled\magstep1             % math extension
\font\twelveit=cmti12                            % italic
\font\twelvesl=cmsl12                            % slanted
\font\twelvebf=cmbx12                            % bold face
\font\twelvett=cmtt12                            % typewriter
\font\twelvess=cmss12                            % sans serif
\font\twelvesc=cmcsc10      scaled\magstep1             % small caps
\font\twelvebi=cmmib10      scaled\magstep1             % bold math

\skewchar\twelvei='177
\skewchar\twelvebi='177
\skewchar\twelvesy='60

\def\twelvepoint{\ifmmode\err@badsizechange\else
	\textfont0=\twelverm \scriptfont0=\eightrm \scriptscriptfont0=\sixrm
	\textfont1=\twelvei  \scriptfont1=\eighti  \scriptscriptfont1=\sixi
	\textfont2=\twelvesy \scriptfont2=\eightsy \scriptscriptfont2=\sixsy
	\textfont3=\twelveex \scriptfont3=\tenex   \scriptscriptfont3=\tenex
	\textfont4=\twelveit \scriptfont4=\eightit \scriptscriptfont4=\sevenit
	\textfont5=\twelvesl
	\textfont6=\twelvebf \scriptfont6=\eightbf \scriptscriptfont6=\sixbf
	\textfont7=\twelvett
	\textfont8=\twelvebi \scriptfont8=\eighti  \scriptscriptfont8=\sixi
	\def\rm{\twelverm\fam=0 }%
	\def\it{\twelveit\fam=4 }%
	\def\sl{\twelvesl\fam=5 }%
	\def\bf{\twelvebf\fam=6 }%
	\def\tt{\twelvett\fam=7 }%
	\def\ss{\twelvess}%
	\def\sc{\twelvesc}%
	\def\bmit{\fam=8 }%
	\rm\setparameters\setbaselines\fi}

%************** 14-point fonts ******************************

\font\fourteenrm=cmr12      scaled\magstep1             % roman
\font\fourteeni=cmmi12      scaled\magstep1             % math italic
\font\fourteensy=cmsy10     scaled\magstep2             % math symbols
\font\fourteenex=cmex10     scaled\magstep2             % math extension
\font\fourteenit=cmti12     scaled\magstep1             % italic
\font\fourteensl=cmsl12     scaled\magstep1             % slanted
\font\fourteenbf=cmbx12     scaled\magstep1             % bold face
\font\fourteentt=cmtt12     scaled\magstep1             % typewriter
\font\fourteenss=cmss12     scaled\magstep1             % sans serif
\font\fourteensc=cmcsc10 scaled\magstep2  % small caps
\font\fourteenbi=cmmib10 scaled\magstep2  % bold math

\skewchar\fourteeni='177
\skewchar\fourteenbi='177
\skewchar\fourteensy='60

\def\fourteenpoint{\ifmmode\err@badsizechange\else
	\textfont0=\fourteenrm \scriptfont0=\tenrm \scriptscriptfont0=\sevenrm
	\textfont1=\fourteeni  \scriptfont1=\teni  \scriptscriptfont1=\seveni
	\textfont2=\fourteensy \scriptfont2=\tensy \scriptscriptfont2=\sevensy
	\textfont3=\fourteenex \scriptfont3=\tenex \scriptscriptfont3=\tenex
	\textfont4=\fourteenit \scriptfont4=\tenit \scriptscriptfont4=\sevenit
	\textfont5=\fourteensl
	\textfont6=\fourteenbf \scriptfont6=\tenbf \scriptscriptfont6=\sevenbf
	\textfont7=\fourteentt
	\textfont8=\fourteenbi \scriptfont8=\tenbi \scriptscriptfont8=\seveni
	\def\rm{\fourteenrm\fam=0 }%
	\def\it{\fourteenit\fam=4 }%
	\def\sl{\fourteensl\fam=5 }%
	\def\bf{\fourteenbf\fam=6 }%
	\def\tt{\fourteentt\fam=7}%
	\def\ss{\fourteenss}%
	\def\sc{\fourteensc}%
	\def\bmit{\fam=8 }%
	\rm\setparameters\setbaselines\fi}

%************** Miscellaneous big fonts *********************

          % roman
             % bold face

%************************************************************
%*
%*            Parameter initialization
%*
%************************************************************

\newdimen\rp@
\newcount\@basestretchnum
\newskip\@baseskip
\newskip\headskip
\newskip\footskip

% Routine to set page parameters

\def\setparameters{\rp@=.1em
	\headskip=24\rp@
	\footskip=\headskip
	\delimitershortfall=5\rp@
	\nulldelimiterspace=1.2\rp@
	\scriptspace=0.5\rp@
	\abovedisplayskip=10\rp@ plus3\rp@ minus5\rp@
	\belowdisplayskip=10\rp@ plus3\rp@ minus5\rp@
	\abovedisplayshortskip=5\rp@ plus2\rp@ minus4\rp@
	\belowdisplayshortskip=10\rp@ plus3\rp@ minus5\rp@
	\normallineskip=\rp@
	\lineskip=\normallineskip
	\normallineskiplimit=0pt
	\lineskiplimit=\normallineskiplimit
	\jot=3\rp@
	\setbox0=\hbox{\the\textfont3 B}\p@renwd=\wd0
	\skip\footins=12\rp@ plus3\rp@ minus3\rp@
	\skip\topins=0pt plus0pt minus0pt}

% Special routine to scale \baselineskip

\def\setbaselines{\maxdepth=4\rp@\baselinestretch=\@basestretchnum}

% The \baselinestretch command

\def\baselinestretch{\afterassignment\@basestretch\@basestretchnum}
\def\@basestretch{%
	\@baseskip=12\rp@ \divide\@baseskip by1000
	\normalbaselineskip=\@basestretchnum\@baseskip
	\baselineskip=\normalbaselineskip
	\bigskipamount=\the\baselineskip
		plus.25\baselineskip minus.25\baselineskip
	\medskipamount=.5\baselineskip
		plus.125\baselineskip minus.125\baselineskip
	\smallskipamount=.25\baselineskip
		plus.0625\baselineskip minus.0625\baselineskip
	\setbox\strutbox=\hbox{\vrule height.708\baselineskip
		depth.292\baselineskip width0pt }}

%************************************************************
%*
%*            Modifications to PLAIN.TEX
%*
%************************************************************

% Modifications to PLAIN routines to handle scaling of page parameters

\def\makeheadline{\vbox to0pt{\baselinestretch=1000
	\vskip-\headskip \vskip1.5pt
	\line{\vbox to\ht\strutbox{}\the\headline}\vss}\nointerlineskip}

\def\makefootline{\baselineskip=\footskip\line{\the\footline}}

\def\big#1{{\hbox{$\left#1\vbox to8.5\rp@ {}\right.\n@space$}}}
\def\Big#1{{\hbox{$\left#1\vbox to11.5\rp@ {}\right.\n@space$}}}
\def\bigg#1{{\hbox{$\left#1\vbox to14.5\rp@ {}\right.\n@space$}}}
\def\Bigg#1{{\hbox{$\left#1\vbox to17.5\rp@ {}\right.\n@space$}}}

% Modifications to PLAIN to handle bold math

\mathchardef\alpha="710B
\mathchardef\beta="710C
\mathchardef\gamma="710D
\mathchardef\delta="710E
\mathchardef\epsilon="710F
\mathchardef\zeta="7110
\mathchardef\eta="7111
\mathchardef\theta="7112
\mathchardef\iota="7113
\mathchardef\kappa="7114
\mathchardef\lambda="7115
\mathchardef\mu="7116
\mathchardef\nu="7117
\mathchardef\xi="7118
\mathchardef\pi="7119
\mathchardef\rho="711A
\mathchardef\sigma="711B
\mathchardef\tau="711C
\mathchardef\upsilon="711D
\mathchardef\phi="711E
\mathchardef\chi="711F
\mathchardef\psi="7120
\mathchardef\omega="7121
\mathchardef\varepsilon="7122
\mathchardef\vartheta="7123
\mathchardef\varpi="7124
\mathchardef\varrho="7125
\mathchardef\varsigma="7126
\mathchardef\varphi="7127
\mathchardef\imath="717B
\mathchardef\jmath="717C
\mathchardef\ell="7160
\mathchardef\wp="717D
\mathchardef\partial="7140
\mathchardef\flat="715B
\mathchardef\natural="715C
\mathchardef\sharp="715D

%************************************************************
%*
%*            Initialization
%*
%************************************************************

\def\err@badsizechange{%
	\immediate\write16{--> Size change not allowed in math mode, ignored}}

\baselinestretch=1000
\tenpoint

\catcode`\@=12                                   % Restore @ sign
% Routine to guarantee that this file is input only once
\catcode`\@=11
\expandafter\ifx\csname @iasmacros\endcsname\relax
	\global\let\@iasmacros=\par
\else  \immediate\write16{}
	\immediate\write16{Warning:}
	\immediate\write16{You have tried to input iasmacros more than once.}
	\immediate\write16{}
	\endinput
\fi
\catcode`\@=12

% Set up font size commands and \baselinestretch command
%\input iasfonts

% Some alternative font names

% Simple spacing commands

\def\doublespace{\baselineskip=2\normalbaselineskip}

% Macros for references and abstracts

\def\nonarrower{\advance\leftskip by-\parindent
	\advance\rightskip by-\parindent}

% Useful commands

\def\boxit#1{\vbox{\hrule\hbox{\vrule\kern3pt
	\vbox{\kern3pt#1\kern3pt}\kern3pt\vrule}\hrule}}

% Special symbols
\def\hence{\leavevmode\hbox{\bf .\raise5.5pt\hbox{.}.} }

\def\dalemb#1#2{{\vbox{\hrule height.#2pt
	\hbox{\vrule width.#2pt height#1pt \kern#1pt \vrule width.#2pt}
	\hrule height.#2pt}}}
\def\gtorder{\mathrel{\raise.3ex\hbox{$>$}\mkern-14mu
	      \lower0.6ex\hbox{$\sim$}}}
\def\ltorder{\mathrel{\raise.3ex\hbox{$<$}\mkern-14mu
	      \lower0.6ex\hbox{$\sim$}}}

% For twoup output
\newdimen\fullhsize
\newbox\leftcolumn
\def\twoup{\hoffset=-.5in \voffset=-.25in
  \hsize=4.75in \fullhsize=10in \vsize=6.9in
  \def\fullline{\hbox to\fullhsize}
  \let\lr=L
  \output={\if L\lr
	 \global\setbox\leftcolumn=\columnbox\global\let\lr=R \advancepageno
      \else \doubleformat \global\let\lr=L\fi
    \ifnum\outputpenalty>-20000 \else\dosupereject\fi}
  \def\doubleformat{\shipout\vbox{
    \fullline{\box\leftcolumn\hfil\columnbox}\advancepageno}}
  \def\columnbox{\leftline{\vbox{\makeheadline\pagebody\makefootline}}}
  \tolerance=1000 }
\twelvepoint
\doublespace

\def\bra#1{{\langle#1|}}
\def\ket#1{{|#1\rangle}}
\def\bracket#1#2{{\langle#1|#2\rangle}}
\def\expect#1{{\langle#1\rangle}}

\def\H{{\hat H}}

\def\Hint{{\hat H}_{\rm int}}

\def\L{{\hat L}}
\def\Ldag{{\hat L}^\dagger}
\def\U{{\hat U}}
\def\Udag{{\hat U}^\dagger}
\def\field{{\hat\phi}}
\def\conj{{\hat\pi}}
\def\P{{\hat{\vec P}}}
\def\OP{{\hat O}}
\def\al{{\hat \alpha}}
\def\be{{\hat \beta}}

\def\QOP{{\hat K}}
\def\SOP{{\hat S}}
\def\F{{\hat F}}
\def\X{{\hat{\vec X}}}

\centerline{\bf Generalized stochastic Schr\"odinger equations}
\centerline{\bf for state vector collapse}

\bigskip

\centerline{Stephen L. Adler and Todd A. Brun}
\centerline{Institute for Advanced Study, Einstein Drive, 
Princeton, NJ 08540}

\bigskip

\centerline{\bf Abstract}

\smallskip

A number of authors have proposed stochastic versions of the Schr\"odinger
equation, either as effective evolution equations for open quantum systems
or as alternative theories with an intrinsic collapse mechanism.  We
discuss here two directions for generalization 
of these equations. First, we study a general class 
of norm preserving stochastic evolution
equations, and show that even after making several specializations, there
is an infinity of possible stochastic Schr\"odinger equations for which
state vector collapse is provable.  Second, we explore the problem of
formulating a relativistic stochastic Schr\"odinger equation,
using a manifestly covariant equation for a quantum field system 
based on the interaction picture of Tomonaga and Schwinger. The stochastic
noise term in this equation can couple to any local scalar density that
commutes with the interaction energy density, and leads to collapse onto
spatially localized eigenstates.  However, as found in a similar model  
by Pearle, the equation predicts an infinite
rate of energy nonconservation proportional to $\delta^3(\vec 0)$, arising 
from the local double commutator in the drift term.  
\vfill\break

\medskip
\centerline{\bf I.~Introduction}
\smallskip

The measurement problem is widely perceived as the greatest difficulty
in the interpretation of quantum mechanics:  how, without invoking a
separate realm of classical measuring devices, can one rule out
superpositions of macroscopically distinct states, as in the famous
Schr\"odinger's cat paradox?  To answer this question, a number of
authors have suggested modifying the usual Schr\"odinger equation
so as to eliminate such superpositions at
large length scales, while retaining the standard quantum results for
microscopic systems.  The result is a modified Schr\"odinger
equation containing extra terms, including {\it stochastic} terms which
reproduce the probabilities of measurements [1--7].

In a parallel development, other researchers have derived effective
equations to describe systems evolving in contact with an external
environment.  These effective equations also take the form of stochastic
Schr\"odinger equations, of a form very similar to those posited in
response to the measurement problem [8--14].

One example of such a modified equation is the {\it quantum state diffusion}
(QSD) equation of Gisin and Percival [12], which has the form
$$
\ket{d\psi} = - i \H \ket\psi dt
  + \sum_k \left( \expect{\Ldag_k}\L_k - {1\over2}\Ldag_k\L_k
  - {1\over2}|\expect{\L_k}|^2 \right) \ket\psi dt
  + \sum_k (\L_k - \expect{\L_k})\ket\psi d\xi_k ~~~. \eqno(1)
$$
Here the {\it Lindblad operators} $\L_k$ [15] represent the effects of the
environment, $\H$ is the Hamiltonian, and the stochastic differentials
$d\xi_k$ represent independent  complex Wiener processes with
vanishing ensemble averages or means (i.e., $M[d\xi_k] = 0$),
that obey the It\^o stochastic calculus
$$
 d\xi_j^* d\xi_k = dt \delta_{jk}~~,~~d\xi_j d\xi_k=dt\, u_{jk}~~,
 ~~dt d\xi_k=0 ~~~. \eqno(2)
$$
Equations (1) and (2) define  an It\^o stochastic differential equation;
in manipulations using the It\^o differential $d$, one must use the
modified chain rule $d(AB)=dAB+A dB+dAdB$.  
[In the ``standard form'' of the QSD equation given by Gisin and Percival, 
the symmetric complex matrix $u_{jk}$ is zero.  However, Wiseman and Di\'osi
have recently shown that the most general It\^o stochastic unraveling of the 
Lindblad evolution has $u_{jk}$ nonzero, with the matrix norm $||u||$ 
bounded by unity [16].  Hence we will keep $u_{jk}$ nonzero in setting up 
the general framework for our discussion, only dropping it later on.]
While the dynamics
of $\ket\psi$ can be extremely complex, there is a tendency for the
state to {\it localize} onto eigenstates of the Lindblad operators
$\L_k$.  Of course, the competing influences of different $\L_k$, or
of the Hamiltonian $\H$, can prevent this localization from taking place.
Note also that Eq.~(1) is nonlinear in $\ket\psi$; this will in general be
necessary for such an equation to preserve the norm of the state.

While we have presented this as an effective equation, arising due to the
effects of an external environment, one can postulate an exactly similar
equation in which the noise is considered fundamental.  Percival has
proposed such an equation with localization onto energy
eigenstates, which he calls {\it Primary} state diffusion (or PSD) [17].
Other such equations have been proposed by Pearle, by Ghirardi, Rimini
and Weber, by Di\'osi, by Ghirardi, Pearle, and Rimini, and by
Hughston [1,3,5,7].
A survey of their properties has recently been given by Adler and Horwitz
[18], who give a detailed discussion of the conditions for the dynamics of
Eq.~(1) to lead to state vector collapse.

Our aim in this paper is twofold.  First, we examine the extent to
which a stochastic dynamics such as Eq.~(1) can be kept in its most
general form, subject to the requirement  that it should still lead to state
vector collapse.  This forms the subject
matter of Sec. II, where we show that there is an infinite parameter
family of stochastic equations for which state vector collapse
is provable.

Our second aim is to explore the well-known problem 
that all equations with the structure of Eq.~(1) are
nonrelativistic.  They are designed to mimic measurement, and they almost
all contain a distinguished frame which takes the role of the rest frame of
the measuring device.  Since in standard QM  measurements take effect
instantaneously on the state vector of the entire system---ultimately,
on the entire universe---it has been very difficult to find a covariant
theory of measurement.
In Sec. III we study a local generalization of
Eq.~(1) which can be written in manifestly covariant form, based on
the ``many fingered time'' Tomonaga-Schwinger generalization of the
Schr\"odinger equation.  (For previous related approaches to this 
problem, see e.g.
[19--23].)  The generalized equation, like its nonrelativistic
counterparts, causes the values of
certain quantities (such as the center of mass of a measuring meter)
to localize.  However, there are difficulties with energy conservation
arising from the local structure of the stochastic terms. 

\medskip
\centerline{\bf II.~Generalized Stochastic Equations}
\smallskip

\medskip
\centerline{\bf II.1~General Framework}
\smallskip

We begin by giving a general framework for the basic QSD
equation of Eq.~(1).   Consider the stochastic state evolution
$$
\ket{d\psi} =\al \ket\psi dt
  + \sum_k \be_k \ket\psi d\xi_k ~~~, \eqno(3)
$$
with $d\xi_k$ independent complex Wiener processes as in Eqs.~(1)
and (2), and with $\al$ and $\be_k$ the operator coefficients
of the drift and stochastic terms respectively, which can also have
an explicit dependence on the state $\ket{\psi}$.  The condition for
the norm of the state to be preserved is
$$
0=d \bra\psi \psi\rangle=\bra{d\psi} \psi \rangle+ \langle \psi \ket{d\psi}
+ \bra{d\psi} d\psi \rangle~~~.\eqno(4)
$$
Substituting Eq.~(3) and its adjoint, and using Eq.~(2) to simplify
the quadratic terms in the It\^o differentials, this becomes
$$
0=dt \bra\psi \al+\al^{\dagger} +\sum_k \be_k^{\dagger} \be_k \ket\psi
+\sum_k[d\xi_k^* \bra\psi \be_k^{\dagger}\ket\psi
+d\xi_k \bra \psi \be_k \ket \psi]~~~.\eqno(5)
$$
Since $d\xi_k$ and $d\xi_k^*$ are independent, Eq.~(5) requires that the
coefficients of $dt$, $d\xi_k$, and $d\xi_k^*$ vanish independently, giving
the conditions
$$
\eqalign{
0=&\bra\psi \left(\al+\al^{\dagger}
  +\sum_k \be_k^{\dagger} \be_k \right)\ket\psi~~~,\cr
0=&\bra\psi \be_k \ket\psi~~~,~~{\rm all}~~ k~~~.\cr
}\eqno(6)
$$
Letting $\L_k$ be a set of general (not necessarily self-adjoint)
operators, and $\H=\H^{\dagger}$ and $\QOP=\QOP^{\dagger}$ be
arbitrary self-adjoint operators,
the general solution to the conditions of Eq.~(6) takes the form
$$
\eqalign{
\be_k=&\L_k-\bra\psi \L_k \ket\psi~~~,\cr
\al=&-i\H +\QOP -\bra\psi \QOP \ket\psi
-{1\over 2} \sum_k \be_k^{\dagger}\be_k~~~,
}\eqno(7)
$$
with the operators $\QOP$ and $\L_k$ still allowed to have an explicit
dependence on the state vector $\ket{\psi}$.
It is convenient for what follows to introduce the definitions
$$\eqalign{
\expect\OP \equiv& \bra\psi \OP \ket \psi~~~,\cr
\Delta \OP \equiv& \OP-\expect\OP~~~,\cr
}\eqno(8)
$$
where $\OP$ is an arbitrary operator.  Then Eq.~(7) can be written in
somewhat more compact form as
$$
\eqalign{
\be_k=&\Delta\L_k~~~,\cr
\al=&-i\H +\Delta\QOP
-{1\over 2} \sum_k \be_k^{\dagger}\be_k~~~.
}\eqno(9)
$$
Equations (3) and (7)-(9) give the general form of a norm-preserving
stochastic extension of the Schr\"odinger equation.  Equation (1)
clearly has this general form, with the specific choice
$\QOP={1\over 2} \sum_k( \expect{\L^{\dagger}_k}\L_k 
- \L^{\dagger}_k\expect{\L_k})$, 
for which $\expect{\QOP}=0$ so $\Delta\QOP=\QOP$.  Usually,
in applications of the QSD equation it is assumed that the Lindblads have
no dependence on the state $\ket{\psi}$, but we will find it useful to
keep open the possibility that they do have a nontrivial state dependence.

To analyze convergence properties implied by this equation, we shall need
formulas for the evolution of the expectation $\expect{\OP}$ and the variance
$V[\OP]\equiv \expect{(\Delta\OP)^2}=\expect{\OP^2}-\expect{\OP}^2$
of a general operator $\OP$.
Using Eq.~(2) and the It\^o extension of the
chain rule, together with Eq.~(3) and its adjoint,
and (in the calculation of $dV$) imposing the normalization
constraints of Eq.~(9), we find after some algebra
the results
$$
\eqalign{
d\expect{\OP}=&\expect{{d\OP\over dt}+\al^{\dagger}\OP+\OP\al
+\sum_k \be^{\dagger}_k \OP \be_k} dt  \cr
+&\sum_k[d\xi_k \expect{\OP \be_k}+d\xi_k^*
\expect{\be^{\dagger}_k \OP}]~~~,\cr
&\cr
dV[\OP]=&[\expect{\Delta\OP {d\OP \over dt}+{d\OP \over dt} \Delta\OP}  \cr
+&\expect{\al^{\dagger} (\Delta \OP)^2 +(\Delta \OP)^2 \al +
\sum_k \be_k^{\dagger} (\Delta \OP)^2 \be_k} \cr
-&2\sum_k \expect{\be_k^{\dagger} \Delta \OP} 
\expect{\Delta \OP \be_k}
 -2 {\rm Re} (\sum_{kl} \expect{\Delta \OP \be_k}
\expect{\Delta \OP \be_l} u_{kl}) 
 ]dt  \cr 
+&\sum_k[ d\xi_k \expect{(\Delta
\OP)^2 \be_k} + d\xi_k^* \expect{\be_k^{\dagger}(\Delta \OP)^2}]~~~.\cr
}\eqno(10)
$$
In applying Eq.~(10), we shall have occasion to take its mean over
the It\^o process.   Since the stochastic
expectation or It\^o process mean $M[~]$ obeys
$$
M[d\xi_k S]=M[d\xi_k^* S] =0~~~,~~{\rm all}~~k~~~,
\eqno(11)
$$
for a general Hilbert space scalar $S$,
the terms in Eq.~(10) involving $d\xi_k$ and $d\xi_k^*$ drop out in
the mean, giving
$$
\eqalign{
M[d\expect{\OP}]=&M[\expect{{d\OP\over dt}+\al^{\dagger}\OP+\OP\al
+\sum_k \be^{\dagger}_k \OP \be_k} ]dt~~~,  \cr
&\cr
M[dV[\OP]]=&M[\expect{\Delta\OP {d\OP \over dt}+{d\OP \over dt} 
\Delta\OP}  \cr
+&\expect{\al^{\dagger} (\Delta \OP)^2 +(\Delta \OP)^2 \al +
\sum_k \be_k^{\dagger} (\Delta \OP)^2 \be_k}
-2\sum_k \expect{\be_k^{\dagger} \Delta \OP}
\expect{\Delta \OP \be_k}\cr
 -&2 {\rm Re} (\sum_{kl} \expect{\Delta \OP \be_k}
\expect{\Delta \OP \be_l} u_{kl}) 
 ]dt~~~.  \cr
}\eqno(12)
$$
Clearly, these equation take the same form if $\OP$ is replaced everywhere by
any function $\F[\OP]$, since this simply defines a new operator $\F$ that
replaces the dummy operator $\OP$.  In the next two sections we shall 
argue that
for the evolution given by Eq.~(3) to converge to an eigenstate of $\OP$,
we must have $M[dV[\OP]] \le 0~~,$
with equality only for $\expect{(\Delta\OP)^2}=0$, and shall demonstrate this
for a particular special class of equations.
\medskip
\centerline{\bf II.2~Specialization}
\smallskip
We shall now introduce some simplifying specializations,
which as we shall see,
still leave an infinite parameter class of stochastic
Schr\"odinger equations, for which state vector reduction to eigenstates
of the operator $\OP$ is provable.  First of all, let us restrict ourselves
to the case in which $\OP$ is a self-adjoint observable, which we assume
to have no explicit time dependence, so that
$\OP=\OP^{\dagger}~~,~~~d\OP/dt=0.$   Secondly, we now take the complex 
matrix $u_{jk}$ of Eq.~(2) to be zero, and we specialize the 
choice of the operators
$\al$ and $\be_k$, which satisfy the
normalization constraints of Eq.~(9), as follows:\hfill \break
\item{(i)}  We take the operator $\QOP$ to be zero, so that
the constraint of Eq.~(6) is satisfied as an operator
relation
$$\al+\al^{\dagger} +\sum_k \be_k^{\dagger} \be_k =0~~~,\eqno(13)$$
which as in Eq.~(9) implies that
$$\al=-i\H -{1\over 2}  \sum_k \be_k^{\dagger} \be_k ~~~.\eqno(14)$$
\hfill\break
\item{(ii)} We take $\H$ to be an operator that commutes with $\OP$,
and take all of the $\L_k$ to be functions solely of the operator $\OP$,
so that they also commute with $\OP$,
$$[\H,\OP]=0~~,~~~\L_k\equiv \L_k[\OP] \Rightarrow [\L_k,\OP]=0
~~~.\eqno(15)$$
Together with Eqs.~(7) and (13), these specializations imply that
$\al$ and $\be_k$ all commute with $\OP$, as well as with any function
$\F[\OP]$ solely of the operator $\OP$,
$$\eqalign{
[\al,\OP]=0~~,~~[\be_k,\OP]=0~~~,\cr
[\al,\F[\OP]]=0~~,~~[\be_k,\F[\OP]]=0~~~,\cr
}\eqno(16)$$

With these specializations, Eqs.~(12) for the time derivatives of the
stochastic mean of the quantum expectation of a function $\F[\OP]$, and
of the stochastic mean of the variance
of $\OP$, simplify dramatically.  Since $\al$  and $\be_k$ commute
with $\OP$, as well as with any function $\F[\OP]$, we have
$$
\eqalign{
\expect{\al^{\dagger} (\Delta \OP)^2
+(\Delta \OP)^2 \al + \sum_k \be_k^{\dagger} (\Delta \OP)^2 \be_k}
=&
\expect{(\Delta \OP)^2 [\al+\al^{\dagger} + \sum_k \be_k^{\dagger} \be_k]}
= 0~~~,\cr
\expect{\al^{\dagger} \F[\OP]+\F[\OP] \al +
\sum_k \be_k^{\dagger} \F[\OP] \be_k}
=&
\expect{\F[\OP] [\al+\al^{\dagger} + \sum_k \be_k^{\dagger} \be_k]}
= 0~~~,\cr
}\eqno(17)
$$
where we have used the operator constraint of Eq.~(13).
Also, since $\expect{\Delta\OP}=0$, we have
$$\expect{ \Delta\OP \be_k}=\expect{\Delta\OP (\L_k - \expect{\L_k})}
=\expect{\Delta\OP \L_k}~~~,\eqno(18)$$
and when $\OP$ is self-adjoint, we have $\expect{\be_k^{\dagger} \Delta 
\OP}=\expect{\Delta \OP \be_k}^*$.
Thus, what remains of Eq.~(12) is
$$
\eqalign{
M[d\expect{\F[\OP]}]=&0~~~,\cr
M[dV[\OP]]=&
-2M[\sum_k|\expect{\Delta\OP \L_k[\OP]}|^2 ]dt ~~~~,  \cr
}\eqno(19)
$$
with $\F[\OP]$ any function solely of the operator $\OP$.
%\vfill\eject
\medskip
\centerline{\bf II.3~State Vector Reduction}
\smallskip
We shall now show that the stochastic dynamics, as specialized in the
preceding subsection, implies state vector reduction to eigenstates of
$\OP$ (assumed nondegenerate), with probabilities given by the Born rule
in terms of the initial wave function.   We shall need one
further assumption beyond
those introduced above, namely, that the scalar valued function $f$ of
$\OP$ defined by
$$
f[\OP]\equiv \sum_k|\expect{\Delta\OP \L_k[\OP]}|^2 ~~~\eqno(20)
$$
vanishes if and only if $\expect{(\Delta\OP)^2}$ vanishes.  One simple way to
achieve this is to take $\L_k[\OP]$ to have the form
$$
\L_k[\OP]=\sum_{n=0}^N c_k^{(n)}(\Delta \OP)^{2n+1}~~~,\eqno(21a)
$$
with $c_k^{(0)}>0$ for at least one value of $k$; note that here we are
using the freedom, remarked on above, to allow the Lindblads to have an
explicit dependence on the state vector.
This implies that for this value of $k$,
$$
\expect{\Delta\OP \L_k[\OP]} =\sum_{n=0}^N c_k^{(n)}
\expect{[(\Delta\OP)^2]^{n+1}}>c_k^{(0)}\expect{(\Delta\OP)^2}
~~~,\eqno(21b)
$$
and so the vanishing of $f[\OP]$ implies the vanishing of
$\expect{(\Delta\OP)^2}$.  This still leaves an infinite parameter freedom
in the construction of the $\L_k$.  A second specific 
example of a $f[\OP]$ with the  needed property is given 
in Sec. III.3 below.  

A general condition for   
$f[\OP]$ to have the needed property can be formulated by 
rewriting Eq.~(20) as 
$$
f[\OP]= \bra{\psi} \Delta\OP  \hat P[\OP] \Delta\OP \ket{\psi}~~,~~~
\hat P[\OP] \equiv \sum_k \L_k[\OP]\ket{\psi} 
\bra{\psi}\L_k^{\dagger}[\OP] ~~~,\eqno(22)
$$
with $\hat P[\OP]$ by construction a positive semidefinite operator.  If the 
Lindblads $\L_k$ were all unity, $\hat P$ would be 
proportional to the projector 
$\ket{\psi}\bra{\psi}$, and since $\ket{\psi}$ is orthogonal to the state 
$\Delta\OP \ket{\psi}$, one would have $f[\OP]\equiv 0$. In order  
for $f[\OP]$ to have the needed property, it is necessary for the Lindblads 
to introduce enough distortion of the projector $\ket{\psi}\bra{\psi}$ 
for $\hat P[\OP]$ to make a strictly positive contribution to Eq.~(22), 
in which  
case the vanishing of $f[\OP]$ requires the vanishing of the state 
$\Delta\OP \ket{\psi}$, or equivalently, the vanishing of 
$\expect{(\Delta \OP)^2}$.  This formulation of the condition on $f[\OP]$  
suggests that in the generic case, it is natural for it to have the needed 
property.  

We can now proceed with a convergence proof, following the presentation
given by Adler and Horwitz [18] (see also [1,4,7]).
Integrating the second line of Eq.~(19) with respect to $t$, we get
$$
M[V[\OP]](t)=M[V[\OP]](0)-2\int_0^t M[f[\OP]](t) dt ~~~.\eqno(23)
$$
Since both $V$ and $f$ are nonnegative, Eq.~(23) implies that the
integrand $M[f[\OP]](t)$ must vanish as $t \to \infty$, since otherwise
the right hand side of Eq.~(23) would become negative at large times.
This in turn implies that $f[\OP](t)$ vanishes as $t \to \infty$ except
on a set of probability measure zero, which by the assumption introduced
following Eq.~(20) implies that the variance $V[\OP]=\expect{(\Delta\OP)^2}$
vanishes as $t \to \infty$ except on a set of probability measure zero.
Thus, when $\OP$ is nondegenerate, the state vector reduces to a pure
state.  Now integrating the first line of Eq.~(19) with respect to $t$,
taking the function $\F[\OP]$ to be a projector $\Pi_{\ell}$ on the
$\ell$th eigenstate of $\OP$, we get
$$
M[\expect{\Pi_{\ell}(\infty)}]=M[\expect{\Pi_{\ell}(0)}]
=\expect{\Pi_{\ell}(0)}
~~~.\eqno(24)
$$
The left hand side of Eq.~(24) is just the probability that the stochastic
process settles at $t=\infty$ on the $\ell$th eigenstate of $\OP$, while
the right hand side of Eq.~(24) is the probability amplitude squared for the
$\ell$th eigenstate to occur in the initial state vector $\ket\psi$.  Thus,
as first proposed by Pearle [1], the first line of Eq.~(19)---which states
that the stochastic process for $\expect{\F[\OP]}$ is a martingale---implies
that the state vector reduction implied by the second line of Eq.~(19)
obeys the Born probability rule.

To compare what we have done to the analyses of Hughston and of Adler
and Horwitz [7,18], those authors consider the energy-driven case in which
the operator $\OP=\H$, and in which one
(at least) of the $\Delta \L_k$ is simply taken as  $\Delta \H$,
corresponding to the case where
the sum in Eq.~(21) consists only of the $n=0$ term.
Note that in this
case, it makes no difference whether we take $\L_k=\H$ or we take
$\L_k=\Delta \H$, since either gives $\Delta\L_k=\Delta\H$.
When $n>0$ terms are present in $\L_k$, this distinction is important,
and plays a role in our example showing that there are more
general stochastic equations that still
allow one to prove state vector reduction.
\medskip
\centerline{\bf III.~Relativistic Stochastic Equations}
\smallskip
Because Eqs.~(1) and (3) involve a universal time variable $t$ at all spatial
points, they are clearly nonrelativistic. This is evident from Eq.~(2),
which states that the same It\^o stochastic differential is present
everywhere in space, giving Wiener processes at space-like separated points
that are totally correlated.  In this section we explore the possibility of
extending Eq.~(1) into an equation with {\it local} Wiener processes, which
can then be generalized to manifestly covariant form.  We shall work
henceforth with a relativistic quantum field theory, rather than with a
nonrelativistic quantum mechanical system, and thus will seek to
modify the Schr\"odinger equation for this field system to a stochastic
Schr\"odinger equation analogous to Eq.~(1), in a manner that
preserves relativistic covariance.  For closely related work, from 
which our analysis differs in some details, see Pearle [19],  Ghirardi, 
Grassi, and Pearle [20], and Di\'osi [23].  
\medskip
\centerline{\bf III.1~The interaction picture}
\smallskip

Suppose we choose a particular Lorentz frame with coordinates $t,\vec x$,
and define a state vector $\ket\psi$ for a field system at
time $t$.  This state evolves to a new state at time $t+dt$ according
to the Schr\"odinger equation
$$
\ket{d\psi(t)} = - i \H \ket{\psi(t)} dt ~~~. \eqno(25)
$$
The Hamiltonian $\H$ is the integral of a Hamiltonian density
$\H(\vec x)$ over a constant time surface,
$$
\H = \int d^3\vec x \, \H(\vec x) ~~~. \eqno(26)
$$
Neither the Hamiltonian $\H$ nor the Hamiltonian density $\H(\vec x)$
are Lorentz invariant, and the Hamiltonian densities at points
$\vec x$ and $\vec y$ do not commute,
$$
[\H(\vec x),\H(\vec y)] = -i{\vec\nabla}_{\vec x} \delta^3(\vec x - \vec y)
  \cdot \left( \P(\vec x) + \P(\vec y) \right) ~~~, \eqno(27)
$$
with $\P(\vec x)$ the momentum density.  These facts make it difficult to
directly extend Eq.~(25) into a stochastic Schr\"odinger equation in a
manner consistent with Lorentz invariance.

As a first step in avoiding these problems, let us switch
to the {\it interaction picture}.  (Our use of this is heuristic and ignores
mathematical issues of the existence of the interaction picture, as discussed
e.g. in [24].)  We write the Hamiltonian density $\H(\vec x)$ as a sum
$$
\H(\vec x) = \H_0(\vec x) + \Hint(\vec x) ~~~, \eqno(28)
$$
where $\H_0(\vec x)$ is the free-field Hamiltonian density and
$\Hint(\vec x)$ is the interaction Hamiltonian density.
Unlike $\H(\vec x)$
as a whole, $\Hint(\vec x)$ is a relativistic invariant in theories
without derivative couplings, and it commutes
with itself at different points,
$$
[\Hint(\vec x),\Hint(\vec y)] = 0 ~~~. \eqno(29)
$$
Let $\U$ be the unitary time evolution operator for the free-field
Hamiltonian,
$$
\U = \exp\{ i \H_0 t \} ~~~. \eqno(30)
$$
If $\ket{\psi_S(t)}$ is the state at time $t$ in the Schr\"odinger picture,
the state in the interaction picture is $\ket{\psi(t)} = \U\ket{\psi_S(t)}$.
We similarly replace the Schr\"odinger picture field operators
(e.g., $\field,\conj$) with interaction picture operators
(e.g., $\field(t) = \U\field\Udag, \conj(t) = \U\conj\Udag$).  If we
express the interaction Hamiltonian density $\Hint(\vec x)$, which is
a function of the field operators at the point $\vec x$, as a function
of the interaction picture field operators, the state then obeys
the simple evolution equation
$$
\ket{d\psi(t)} = - i \Hint \ket{\psi(t)} dt ~~~, \eqno(31)
$$
where
$$
\Hint = \int d^3\vec x \, \Hint(\vec x) ~~~. \eqno(32)
$$
One must now remember that operators that were time independent in the 
Schr\"odinger picture acquire a time dependence governed  
by the free Hamiltonian $\H_0$.

So far this discussion has been restricted to constant-time surfaces
in a single Lorentz frame, in which a fixed time step $dt$ is taken
simultaneously at all spatial points $\vec x$, and so Eq.~(31) is
still not Lorentz invariant.    We
now follow Tomonaga and Schwinger [25,26] (see also Matthews, Kroll, and 
Dyson [27--29])
in generalizing Eq.~(32) into a
{\it local} evolution equation.
Consider a spacelike surface $\sigma$ with local coordinates $\vec x$,
on which the state of the underlying quantum fields 
is described by a Fock space state vector $\ket{\psi(\sigma)}$.
Instead of advancing the whole spacelike surface $\sigma$, we instead
move the surface forward (i.e., in the normal direction) by an increment 
$dt(\vec x)$ only in the vicinity
of a single point $\vec x$, distorting the surface $\sigma$ to a new
spacelike surface $\sigma^{\prime}$.
Under this evolution, the state vector $\ket\psi$ evolves to a
state vector $\ket\psi +d_{\vec x}\ket\psi$, with the 
change in the state vector given by 
$$
d_{\vec x}\ket\psi = - i \Hint(\vec x) \ket\psi dt(\vec x) ~~~.\eqno(33)
$$
The change in the state vector resulting from advancing the {\it entire}
surface is then
$$
\ket{d\psi} = \int d^3\vec x \, d_{\vec x}\ket\psi ~~~.\eqno(34)
$$
Since the $\Hint(\vec x)$ at all points commute, the order in which the
spacelike surface is advanced is immaterial and so the right hand side of
Eq.~(34) can be unambiguously integrated, which for constant-time 
surfaces with $dt(\vec x)\equiv dt$ 
recovers the original interaction picture Schr\"odinger equation
of Eqs.~(31)-(32).  From the viewpoint of constructing a stochastic
generalization, the local form of the interaction picture
evolution equation given in Eq.~(33)
has three advantages:  It is readily put in manifestly covariant form, it
involves only the Lorentz scalar operator density $\Hint(\vec x)$, and this
operator commutes with itself (and with other easily constructed
scalar densities) at spacelike separations.

\medskip
\centerline{\bf III.2~The local norm-preserving stochastic equation}
\smallskip

Let us now replace the local unitary evolution equation of Eq.
~(33) with a new equation
$$
d_{\vec x} \ket\psi =
\al(\vec x) \ket\psi dt(\vec x) + \be(\vec x) \ket\psi d\xi(\vec x)
~~~,\eqno(35)
$$
in which we take the coefficient functions $\al(\vec x)~,~\be(\vec x)$ and
$\al(\vec y)~,~\be(\vec y)$  to mutually commute for all $\vec x~,~\vec y$,
so that no noncommutativity problems are encountered when we compound
evolutions for different values of $\vec x$.
Here $d\xi(\vec x)$ is a complex stochastic differential variable
defined at each point $\vec x$, which has zero stochastic mean
(i.e., $M[d\xi(\vec x)] = 0$), and which obeys the local It\^o calculus
$$
d\xi^*(\vec x) d\xi(\vec y) = \delta^3(\vec x - \vec y) dt(\vec x)~~,~~
d\xi(\vec x) d\xi(\vec y)=dt(\vec x) d\xi(\vec y)=0~~~.\eqno(36)
$$
The spatially integrated form corresponding to Eq.~(35) is
$$
\ket{d\psi} = \int d^3\vec x \, d_{\vec x}\ket\psi
=\int d^3\vec x [\al(\vec x) \ket\psi dt(\vec x)
+ \be(\vec x) \ket\psi d\xi(\vec x)]
~~~.\eqno(37)
$$
In analogy with our discussion of Sec. II.1, we can now determine
the conditions on the coefficient functions $\al(\vec x)$ and $\be(\vec x)$
for Eq.~(37) to preserve the norm of the state,
$$
\eqalign{
d\bracket\psi\psi =& \bracket{d\psi}\psi + \bracket\psi{d\psi}
  + \bracket{d\psi}{d\psi} \cr
=& \int d^3\vec x \,
\bra\psi [ \al(\vec x) +\al(\vec x)^{\dagger}
+\be(\vec x)^{\dagger} \be(\vec x)  ] \ket\psi dt(\vec x)   \cr
+& \int d^3\vec x \, [\bra\psi \be(\vec x)^{\dagger} \ket\psi d\xi^*(\vec x)
+\bra\psi \be(\vec x) \ket\psi d\xi(\vec x) ] ~~~.\cr
}\eqno(38)
$$
Since $d\xi^*(\vec x)$, $d\xi(\vec x)$, and $dt(\vec x)$ are linearly
independent,  the normalization of the state $\bracket\psi\psi=1$
is preserved if and only if for all $\vec x$ we impose the conditions
$$
\eqalign{
0=&\bra\psi [ \al(\vec x) +\al(\vec x)^{\dagger}
+\be(\vec x)^{\dagger} \be(\vec x)  ] \ket\psi~~~,\cr
0=&\bra\psi \be(\vec x) \ket\psi  ~~~.\cr
}\eqno(39)
$$
Evidently, if we were to  replace $\al$ in Sec. II.1 by $\sum_k \al_k$,
then Eq.~(39) could be viewed as a local version of Eq.~(6), with $\vec x$
playing the role of the index $k$.   Imposing the normalization 
conditions, and specializing henceforth to $dt(\vec x)\equiv dt$ and flat 
spacelike surfaces $\sigma$, 
we find the following local version of Eq.~(10),
$$
\eqalign{
d\expect{\OP}=&\expect{d\OP}+\int d^3 \vec x \,
[\expect{\al^{\dagger}(\vec x)\OP+\OP\al(\vec x)
+ \be^{\dagger}(\vec x) \OP \be(\vec x)} ]dt  \cr
+&\int d^3 \vec x [d\xi(\vec x) \expect{\OP \be(\vec x)}+d\xi^*(\vec x)
\expect{\be^{\dagger}(\vec x) \OP}]~~~,\cr
&\cr
dV[\OP]=&\expect{\Delta \OP d\OP+d\OP \Delta \OP}\cr
 +&\int d^3\vec x \,
[\expect{\al^{\dagger}(\vec x) (\Delta \OP)^2 +(\Delta \OP)^2 \al(\vec x)
+ \be^{\dagger}(\vec x) (\Delta \OP)^2 \be(\vec x)}
-2\expect{\be^{\dagger}(\vec x) \Delta \OP}
\expect{\Delta \OP \be(\vec x)} ]dt  \cr
+&\int d^3\vec x \,
[ d\xi(\vec x) \expect{(\Delta \OP)^2 \be(\vec x)}
+ d\xi^*(\vec x) \expect{\be^{\dagger}(\vec x)(\Delta \OP)^2}]~~~.\cr
}\eqno(40)
$$

Instead of working with the most general form of the normalization condition,
we shall specialize (as we did in Sec. II.2) and satisfy Eq.~(39) by
taking $\al(\vec x)$ and $\be(\vec x)$ to have
the form
$$
\eqalign{
\al(\vec x)=&-i\Hint(\vec x)
-{1 \over 2} \be(\vec x)^{\dagger} \be(\vec x)~~~,\cr
\be(\vec x)=&\Delta \SOP(\vec x)~~~,\cr
}\eqno(41)
$$
with $\SOP(\vec x)$ any local Lorentz scalar operator that commutes with
$\Hint(\vec x)$.  We shall further assume $\SOP(\vec x)$  to be
self-adjoint.  Additionally, we shall assume that
the operator $\OP$ is self-adjoint and has no intrinsic time 
dependence in the Schr\"odinger
picture, so that in the interaction picture its time dependence is given by
$${d \OP \over dt}=i[\H_0,\OP]~~~.\eqno(42)$$
With these specializations, Eq.~(40) can be rewritten after a little
algebra as
$$
\eqalign{
d\expect{\OP}=&
\left[\expect{i[\H,\OP] -{1\over 2}\int d^3 \vec x \, 
[\SOP(\vec x),[\SOP(\vec x),\OP]]} \right]dt
\cr
+&\int d^3 \vec x [d\xi(\vec x) \expect{\OP \Delta\SOP(\vec x)}
+d\xi^*(\vec x)\expect{\Delta\SOP(\vec x) \OP}]
 ~~~,\cr
&\cr
dV[\OP]=&
\left[\expect{i[\H,(\Delta \OP)^2]-{1\over 2}\int d^3\vec x \,
[\SOP(\vec x),[\SOP(\vec x),(\Delta\OP)^2]]}
-2\int d^3\vec x \, |\expect{\Delta \OP \Delta\SOP(\vec x)}|^2 \right]
dt  \cr
+&\int d^3\vec x \,
[d\xi(\vec x)\expect{(\Delta \OP)^2 \Delta\SOP(\vec x)}+
d\xi^*(\vec x)\expect{ \Delta\SOP(\vec x)(\Delta \OP)^2}]
~~~.\cr
}\eqno(43)
$$
Although not needed for our purposes, by using the fact that 
$d^3\vec x  dt$ and $\SOP(\vec x)$ are Lorentz scalars, the 
stochastic and drift terms in Eq.~(43) can be readily written 
in manifestly covariant form.  The 
corresponding covariant transcription of the Hamiltonian evolution 
terms is given in Matthews [27] and Kroll [28].  

\medskip
\centerline{\bf III.3~Reduction for local density eigenstates}
\smallskip
Let us now apply the above formulas to
discuss state vector reduction to local density eigenstates, giving
a relativistic generalization of the localization models discussed in
 [3--5].  Let us make the specific choice
$$
\SOP(\vec x) =C\Hint(\vec x)~~~,\eqno(44)
$$
which obviously satisfies the commutativity conditions
$$
[\al(\vec x),\al(\vec y)]=[\al(\vec x),\be(\vec y)]=
[\be(\vec x),\be(\vec y)]=[\al(\vec x),\Hint(\vec y)]=[\be(\vec x),\Hint(\vec y)]
=0~~~,\eqno(45)
$$
for all spacelike separated points $\vec x~,~\vec y$.  In field theories
like the Standard Model, in which all mass comes from spontaneous symmetry
breaking, the mass terms arise from $\Hint(\vec x)$, and so for bulk matter
we are effectively taking  $\SOP$ to be the local mass density operator,
multiplied by a scale factor $C$.

As a concrete illustration of how Eq.~(43) can lead to state vector
reduction and localization, let us consider the simplified case of an
apparatus connected to a pointer with two macroscopic states specified
by two values $\vec X_1~,~~\vec X_2$ of the pointer center of mass variable
$\X$,
$$\X\equiv {\int_{\rm pointer} d^3\vec x \vec x \SOP(\vec x) \over
  \int_{\rm pointer} d^3\vec x  \SOP(\vec x)}~~~.\eqno(46)$$
We shall apply Eq.~(43) to this system, taking $\OP=\X$.  By Eq.~(45),
the double commutators $[\SOP(\vec x),[\SOP(\vec x),\OP]]$ and
$[\SOP(\vec x),[\SOP(\vec x),(\Delta\OP)^2]]$  both vanish, but in general
the commutators $[\H,\OP]$ and  $[\H,(\Delta \OP)^2]$ are nonzero.
However, if we take the two macroscopic pointer positions to be degenerate
in energy, then the commutators involving $\H$ vanish within the
degenerate two-state subspace.  Taking the stochastic mean
$M[~]$ of Eqs.~(43), we then find within the two-state subspace the
simplified equations
$$
\eqalign{
M[d\expect{\X}]=&0 ~~~,\cr
M[dV[\X]]=&-2\int d^3\vec x
M[|\expect{\Delta \X \Delta \SOP (\vec x)}|^2] dt  ~~~.\cr
}\eqno(47)$$
We have here exactly the same structure as we found in Eq.~(19) above, and
the function $f(\X)\equiv \int d^3 \vec x 
|\expect{\Delta \X \Delta \SOP (\vec x)}|^2$ is easily seen [c.f. the final 
line in Eq.~(48) below]  to obey the condition that the vanishing of $f(\X)$ 
implies the vanishing of $\expect{(\Delta \X)^2}$.  
Hence the same argument as was used in Eqs.~(23) and (24) proves that an
initial superposition of the two center of mass eigenstates reduces to
either the state with $\X=\vec X_1$ or the state with $\X=\vec X_2$, with 
respective
probabilities given by the amplitude squared to find the initial state in
the respective $\X$ eigenstate.

  From Eq.~(47), we can estimate the reduction rate $\Gamma$ as follows.
Writing $\ket{\psi}=\ket{\vec X_1} \cos \theta + \ket{\vec X_2}\sin 
\theta$, and assuming that the states $\ket{\vec X_1}~,~\ket{\vec X_2}$ 
differ sufficiently for us to approximate that $\bra{\vec X_1}
\SOP(\vec x)\ket{\vec X_2} \simeq 0$, 
we have after a short calculation
$$
\eqalign{
V[\X]=&\expect{(\Delta \X)^2}=
 \sin^2\theta \cos^2 \theta (\vec X_1-\vec X_2)^2~~~,\cr
\expect{\Delta \X \Delta \SOP (\vec x)} =&
 \sin^2\theta \cos^2\theta  (\vec X_1-\vec X_2)
(\bra{\vec X_1} \SOP(\vec x) \ket{\vec X_1}
-\bra{\vec X_2} \SOP(\vec x) \ket{\vec X_2})~~~,\cr
|\expect{\Delta \X \Delta \SOP (\vec x)}|^2 =&
 \sin^4\theta \cos^4 \theta (\vec X_1-\vec X_2)^2
|\bra{\vec X_1} \SOP(\vec x) \ket{\vec X_1}
-\bra{\vec X_2} \SOP(\vec x) \ket{\vec X_2}|^2\cr
=&\expect{(\Delta \X)^2}^2 \, 
{|\bra{\vec X_1} \SOP(\vec x) \ket{\vec X_1}
-\bra{\vec X_2} \SOP(\vec x) \ket{\vec X_2}|^2  
\over  (\vec X_1-\vec X_2)^2}~~~.\cr
}\eqno(48)$$
Thus Eq.~(47) becomes
$$
{M[ d(\sin^2 \theta \cos^2 \theta)]\over dt} =
-2\int d^3\vec x \,  M[ \sin^4\theta \cos^4 \theta ]
|\bra{\vec X_1} \SOP(\vec x) \ket{\vec X_1}
-\bra{\vec X_2} \SOP(\vec x) \ket{\vec X_2}|^2~~~,\eqno(49)
$$
from which we see that, up to numerical factors of order unity, the
reduction rate is given by
$$
\eqalign{
\Gamma \sim&
\int d^3\vec x \,  |\bra{\vec X_1} \SOP(\vec x) \ket{\vec X_1}
-\bra{\vec X_2} \SOP(\vec x) \ket{\vec X_2}|^2\cr
\sim&C^2 \int_{\rm pointer}d^3\vec x \, [{\rm ~Mass~Density}]^2~~~.\cr
}\eqno(50)
$$
  For a pointer containing $N\sim 10^{23}$ nucleons of
mass $M\sim 1~ {\rm GeV}$ and volume  $V\sim 10^{-39} ~{\rm cm}^3$, the
estimate of Eq.~(50) becomes
$$\Gamma \sim C^2 N M^2V^{-1}~~~,\eqno(51)$$
which gives a reduction rate $\Gamma > 10^8~ {\rm sec}^{-1}$  (corresponding  
to a collapse time faster than characteristic observational time scales)   
for $C > (10^{9}~ {\rm GeV})^{-2}$.  This corresponds to a mass scale at
roughly the geometric mean between the Planck mass and a nucleon mass.
Thus, in contrast to the energy driven model [7,12,18] for state 
vector reduction,
where the mass scale for the
coefficient of the noise terms is Planckian, in the local version discussed
here the mass scale for the noise terms is much below the Planck scale,
but still large compared to elementary particle masses.
\medskip
\centerline{\bf III.4~Energy nonconservation}
\smallskip
Except for the special case of stochastic equations in which the
Lindblads are taken to be operators that commute with the Hamiltonian
(including the Hamiltonian itself), stochastic modifications of the
Schr\"odinger equation lead to energy nonconservation, as has been noted
in the papers of Ghirardi, Rimini, and Weber, and of Pearle [1,3,19].
Let us examine this issue in the context of the relativistic model 
discussed above.  For any operator
$\OP$, the stochastic expectation $M[~]$ of the first formula in Eq.~(43) is
$$
M[d\expect{\OP}]=
M\left[\expect{i[\H,\OP] -{1\over 2}\int d^3\vec x  
[\SOP(\vec x),[\SOP(\vec x),\OP]]} \right]dt 
~~~,\eqno(52)
$$
which when applied to the Hamiltonian (i.e., taking $\OP=\H$) gives
for the mean rate of energy nonconservation
$$
M\left[{d\expect{\H}\over dt}\right]= -{1\over 2} \int d^3\vec x \,
M[\expect{ [\SOP(\vec x),[\SOP(\vec x),\H]]} ]
~~~.\eqno(53)
$$

In typical field theory models, the double commutator appearing
in Eq.~(53) is
not only nonzero, but as first noted by Pearle [19]  is 
proportional to $\delta^3(\vec 0)$ and thus
is infinite.   For example, taking a Dirac field model with
$$
\H=\int d^3\vec x \, {\hat\psi}^{\dagger}(\vec x)[i^{-1}\vec \alpha \cdot
\vec \nabla + \beta \phi(\vec x)] \hat \psi(\vec x)=\H_0+\Hint~~~,\eqno(54)
$$
with $\phi(\vec x)$ an external scalar field with nonzero vacuum expectation,
and choosing 
$$\SOP(\vec x)=C \Hint(\vec x)=C\hat\psi^{\dagger}(\vec x) \beta \phi(\vec x)
\hat\psi(\vec x)~~~,\eqno(55)
$$
one has
$$
[\SOP(\vec x),[\SOP(\vec x),\H]] =
\delta^3(\vec 0)C^2 \phi^2(\vec x) i^{-1}[{\hat \psi}^{\dagger}(\vec x)
\vec \alpha \cdot \vec \nabla \hat \psi(\vec x)
- \vec \nabla {\hat \psi}^{\dagger}(\vec x) \cdot \vec
\alpha \hat  \psi(\vec x)]~~~.\eqno(56)
$$
Similar results are found in scalar meson field theory models, and 
appear to be generic.  Moreover, except for special choices of 
$\SOP(\vec x)$ (see, e.g. [19-21]), the coefficient of $\delta^3(\vec 0)$ 
is a nontrivial operator and not a constant.  The $\delta^3(\vec 0)$ 
singularity is a direct result of the local derivative structure of the drift term, and we have not found a mechanism to cancel it within the 
standard stochastic differential equation and quantum field theory 
framework discussed here.

\medskip
\centerline{\bf IV.~Conclusions}
\smallskip

We have presented two generalizations of stochastic Schr\"odinger equations
for state vector collapse.  First, we have shown that there is an infinite
parameter family of such equations for which one can prove
state vector collapse with probabilities given by the Born rule.
Second, we have given a relativistic stochastic equation which can be made
manifestly covariant, and which produces localization onto mass density
eigenstates.  This produces spatial localization  for superpositions
of macroscopically distinct system states; to give rapid enough
state vector localization in plausible experimental setups, the scale
mass governing the stochastic terms must be considerably smaller than the
Planck mass.  The local equation has the defect that it 
leads to a divergent rate of energy
nonconservation in generic field theory models, indicating that 
new ideas will be needed to achieve a satisfactory relativistic state 
vector collapse model.  

\medskip
\centerline{\bf Acknowledgments}
\smallskip

We gratefully acknowledge useful conversations with Ian Percival, and email 
correspondence with L. Di\'osi and M. Wiseman.  
The authors also acknowledge support by DOE Grant No. DE-FG02-90ER40542.

\vfill\eject

\centerline{\bf References}
\smallskip

[1] P. Pearle, Phys. Rev. D {\bf 13}, 857 (1976);
Int. J. Theor. Phys. {\bf 18}, 489 (1979).
\smallskip

[2] N. Gisin, Phys. Rev. Lett. {\bf 52}, 1657 (1984).
\smallskip

[3] G.C. Ghirardi, A. Rimini and T. Weber, Phys. Rev. D {\bf 34}, 470 (1986).
\smallskip

[4]  G.C. Ghirardi, P. Pearle, and A. Rimini, Phys. Rev. A {\bf 42}, 78
(1990).
\smallskip

[5] L. Di\'osi, J. Phys. A {\bf 21}, 2885 (1988);
Phys. Lett. {\bf 129}A, 419 (1988); Phys. Lett.
{\bf 132}A, 233 (1988).
\smallskip

[6] N. Gisin, Helv. Phys. Acta. {\bf 62}, 363 (1989).
\smallskip

[7] L.P. Hughston, Proc. Roy. Soc. London A{\bf 452}, 953 (1996).
\smallskip

[8] A. Barchielli and V.P. Belavkin, J. Phys. A {\bf 24}, 1495 (1991).
\smallskip

[9] J. Dalibard, Y. Castin and K. M\o lmer, Phys. Rev. Lett. {\bf 68},
580 (1992).
\smallskip

[10] R. Dum, P. Zoller and H. Ritsch, Phys. Rev. A {\bf 45}, 4879 (1992).
\smallskip

[11] C.W. Gardiner, A.S. Parkins and P. Zoller, Phys. Rev. A {\bf 46},
4363 (1992).
\smallskip

[12] N. Gisin and I.C. Percival, J. Phys. A {\bf 25}, 5677 (1992);
J. Phys. A {\bf 26}, 2233, 2245 (1993).  
\smallskip

[13] H.J. Carmichael, {\sl An Open Systems Approach to Quantum Optics},
(Springer, Berlin, 1993).
\smallskip

[14] P. Goetsch and R. Graham, Ann. Phys., Lpz. {\bf 2}, 706 (1993).
\smallskip

[15] G. Lindblad, Commun.~Math.~Phys.~{\bf 48}, 119 (1976).
\smallskip

[16]  H. M. Wiseman and L. Di\'osi, ``Complete parameterization, and 
invariance, of diffusive quantum trajectories for Markovian open systems'', 
quant-ph/0012016.  
\smallskip

[17] I.C. Percival, Proc.~Roy.~Soc.~A {\bf 447}, 189 (1994);
{\it ibid.\/} {\bf 451}, 503 (1995).
\smallskip

[18] S.L. Adler and L.P. Horowitz, J. Math. Phys. {\bf 41}, 2485 (2000).
\smallskip

[19]  P. Pearle, ``Relativistic Collapse Model With Tachyonic Features,''
Hamilton College preprint;
``Collapse Models,'' in H.-P. Breuer and F. Petruccione,
eds., Ref.~[21];
``Relativistic Model for Statevector Reduction'', in
P. Cvitanovi\'c, I. Percival, and A. Wirzba, eds., {\sl Quantum
Chaos---Quantum Measurement} (Kluwer, Dordrecht, 1991); ``Toward a 
Relativistic Theory of Statevector Reduction,'' in A. I. Miller, ed., 
{\sl Sixty-Two Years of Uncertainty: Historical, Philosophical, 
and Physical Inquiries into the Foundations of Quantum Mechanics} 
(Plenum Press, 
New York, 1990).  
\smallskip

[20] G.C. Ghirardi, ``Some Lessons from Relativistic Reduction Models'',
in H.-P. Breuer and F. Petruccione, eds., Ref. [21].
\smallskip

[21] G. Ghirardi, R. Grassi, and P. Pearle, ``Relativistic Dynamical
Reduction Models and Nonlocality'', in P. Lahti and P. Mittelstaedt,
eds., {\sl Symposium on the Foundations of Modern Physics, 1990:
Quantum Theory of Measurement and Related Philosophical Problems}
(World Scientific, Singapore, 1991).
\smallskip

[22]  H.-P. Breuer and F. Petruccione, J. Phys. A {\bf 31}, 33 (1998),
and ``State Vector Reduction in Relativistic Quantum Mechanics: an
Introduction'', in H.-P. Breuer and F. Petruccione, eds., {\sl Open
systems and measurements in relativistic quantum theory}, Lecture Notes in
Physics V. 526 (Springer, Berlin, 1999).
\smallskip

[23]  L. Di\'osi, Phys. Rev. A {\bf 42}, 5086 (1990); L. Di\'osi, J. Phys. 
A:  Math. Gen. {\bf 31}, 9601 (1998).  
\smallskip

[24] R.F. Streater and A.S. Wightman, {\sl PCT, Spin and Statistics,
and All That}, (Benjamin, New York, 1964).
\smallskip

[25] S. Tomonaga, Prog. Theor. Phys. {\bf 1}, 27 (1946).
\smallskip

[26] J. Schwinger, Phys. Rev. {\bf 74}, 1492 (1948).

[27] P.T. Matthews, Phys. Rev. {\bf 75}, 1270 (1949).
\smallskip

[28] N.M. Kroll, Phys. Rev. {\bf 75}, 1321A (1949).
\smallskip

[29] F.J. Dyson, Phys. Rev. {\bf 75}, 486 (1949).
\smallskip

\smallskip
\vfill
\bye